\documentclass[12pt]{article}
\usepackage{epsf,amsfonts,amssymb,epsfig,amsmath,graphics,subfig}
\usepackage{hyperref,graphicx, subfig}
\renewcommand{\text}[1]{#1}

\newcommand{\be}{\begin{equation}}
\newcommand{\ee}{\end{equation}}
\newcommand{\ben}{\begin{displaymath}}
\newcommand{\een}{\end{displaymath}}
\newcommand{\bea}{\begin{eqnarray}}
\newcommand{\eea}{\end{eqnarray}}
\newcommand{\ba}{\begin{align}}
\newcommand{\ea}{\end{align}}
\newcommand{\nn}{\nonumber \\}
\newcommand{\bi}{\begin{itemize}}
\newcommand{\ei}{\end{itemize}}

\begin{document}

\makeatletter
\renewcommand{\theequation}{\thesection.\arabic{equation}}
\@addtoreset{equation}{section}
\makeatother

\baselineskip 18pt

\begin{titlepage}

\vfill

\begin{flushright}
Imperial/TP/2011/JG/02\\
\end{flushright}

\vfill

\begin{center}
   \baselineskip=16pt
   {\Large\bf  Holographic striped phases}
  \vskip 1.5cm
      Aristomenis Donos and Jerome P. Gauntlett\\
   \vskip .6cm
      \begin{small}
      \textit{Blackett Laboratory, 
        Imperial College\\ London, SW7 2AZ, U.K.}
        %E-mail: j.gauntlett, d.waldram@imperial.ac.uk}
        \end{small}\\*[.6cm]
 %  \vskip .3cm
   %   \begin{small}
     % \textit{$^2$Perimeter Institute for Theoretical Physics\\
        %Waterloo, Ontario, N2L 2Y5, Canada}
        %E-mail: j.gauntlett, d.waldram@imperial.ac.uk}
        %\end{small}\\*[.6cm]

\end{center}

\vfill

\begin{center}
\textbf{Abstract}
\end{center}

\begin{quote}
We discuss new types of instabilities of $D=4$ electrically charged AdS-Reissner-Nordstr\"om
black branes that involve neutral pseudo-scalars.  
The instabilities spontaneously break translational invariance
and
are associated with the dual three-dimensional CFTs, at finite temperature and fixed chemical potential
with respect to a global abelian symmetry, acquiring
striped phases. We show that such instabilities are present for the infinite class of skew-whiffed $AdS_4\times SE_7$
solutions of $D=11$ supergravity, albeit at a lower temperature than the known superfluid instabilities.
\end{quote}

\vfill

\end{titlepage}
\setcounter{equation}{0}

%%%%%%%%%%%%%%%%%%%%%%%%%%%%%%%%%%%%%%%%%%%%%%%%%%%%%%%%%%%%%%%%%%%%%%%
%\tableofcontents
%%%%%%%%%%%%%%%%%%%%%%%%%%%%%

%\begin{thebibliography}{99}
\section{Introduction}

The AdS/CFT correspondence provides an important framework for studying the behaviour of strongly coupled
quantum field theories. Recently the possible applications to condensed matter physics have received particular attention.
One focus has been on the phase structure of conformal field theories when held at non-zero chemical potential with respect to a global abelian symmetry. 
%The bulk gravitational description includes 
%a metric and a $U(1)$ gauge field as well as other possible fields.
At high temperatures the CFTs are described by electrically charged 
AdS-Reissner-Nordstr\"om (AdS-RN)  black branes. As the temperature is lowered
these black branes can become unstable giving rise to new black brane solutions which then describe
new phases of the dual field theory.

A well studied class of instabilities of the electrically charged  AdS-RN black branes lead to the
spontaneous breaking of the
global abelian symmetry and hence to superfluid phases. 
Such instabilities can occur when the metric and gauge field are coupled to charged fields and the superfluid phases 
are described by electrically charged AdS black branes with additional charged hair. 
These instabilities and the back reacted black branes
were first studied in a bottom up context in
\cite{Gubser:2008px}\cite{Hartnoll:2008vx}\cite{Hartnoll:2008kx}
and then subsequently embedded into D=11 and D=10 
supergravity in \cite{Denef:2009tp}\cite{Gauntlett:2009dn}\cite{Gauntlett:2009bh}\cite{Gubser:2009qm}. They were
studied using D-brane probes in \cite{Ammon:2008fc}.

In this paper we will present a new class of instabilities of electrically charged AdS-RN black branes in $D=4$
which lead to phases that spontaneously break the spatial translation invariance of the dual field theory. 
Such phases appear in condensed matter physics in a variety of settings. 
More precisely, the phases spontaneously break some or all of the symmetries of the underlying lattice.
Examples include charged density wave (CDW) phases and spin density wave (SDW) phases 
which involve, as
the names suggest, a spatial modulation of the charge density and the spin density, respectively (see
\cite{Gruner:1988zz}\cite{Gruner:1994zz} for reviews).
Other orders include staggered flux phases \cite{sf} and more general density waves with non-zero angular momentum 
\cite{nay}. These orders have been discussed in the context of both heavy fermion and the cuprate superconductors.
In particular, it is well established that the cuprate superconductors are characterised by a rich set of competing orders.
The undoped materials are antiferromagnetic Mott insulators, but doping leads to superconductivity and
striped phases, in which there is unidirectional charge and spin density waves.
There are also additional
ordered phases, including nematic phases and circulating current phases which do not break the translational
invariance of the lattice.
A review of this area can be found in \cite{vojta} and a historical overview can be found
in \cite{Zaanen:2010yk}. 

The new instabilities of the AdS-RN black branes that we discuss here
will lead to black brane solutions that are holographically dual to striped phases.
More precisely, near the temperature $T_c$ at which 
these ``striped black branes" appear, the 
 $d=3$ current of the global symmetry in the CFT, 
dual to the bulk gauge field, spontaneously acquires a spatially modulated vev of the form
\begin{align}\label{eqnone}
\langle j_t\rangle -\bar j_t \propto\cos(2k_c x),\qquad
\langle j_x\rangle = 0\qquad
\langle j_y\rangle\propto\sin(k_c x)
\end{align}
where $\bar j_t$ is a spontaneously generated constant component.
Note that the translation invariance in the $x$ direction is spontaneously broken but
it is preserved in the $y$ direction. These stripes combine a CDW (corresponding to the spatially modulated part of
$ j_t$) with a ``current density wave"\footnote{If one gauges the
global symmetry in the boundary field theory, these currents would give rise to a spatially modulated 
magnetic field.} (corresponding to the spatial modulation of
$ j_x,j_y$). As one moves away from $T_c$ higher harmonics will
appear, but the spatial modulation of the CDW will have a period that is half of the
current density wave.

The simplest setting in which these striped instabilities can occur is
when the metric and gauge field are coupled to a neutral pseudo-scalar $\varphi$. Indeed a key coupling
driving the instability is the coupling $\varphi F\wedge F$, where $F$ is the field strength of the abelian
gauge field. In the context of Kaluza-Klein reductions of $D=10$ and $D=11$ supergravity these types of couplings
are commonplace and so we expect that the instabilities that we discuss, and straightforward generalisations thereof,
will be widespread.  Here we will show that they are present in the context of the infinite class of
skew-whiffed $AdS_4\times SE_7$ solutions of $D=11$ supergravity, where $SE_7$ is a seven-dimensional
Sasaki-Einstein space. It has been shown in \cite{Gauntlett:2009zw}\cite{Gauntlett:2009dn}\cite{Gauntlett:2009bh}
that there is
a consistent Kaluza-Klein truncation of $D=11$ supergravity 
on an arbitrary $SE_7$ to a $D=4$ theory of gravity involving a metric, a
gauge field, a neutral pseudo-scalar and a charged scalar. We show that this model exhibits spatially modulated instabilities,
with vanishing charged scalar, and we determine the highest temperature at which they can occur. Although this demonstrates that such instabilities
are indeed present in a top down context, it should be noted that for this particular class of theories,
the known superfluid instability involving the 
charged scalar field, found in \cite{Gauntlett:2009dn}\cite{Gauntlett:2009bh}, already sets in at a higher temperature.

We will also consider a more general class of models which couple 
the metric, a gauge field and neutral pseudo-scalar to
an additional massive vector field. These models naturally appear in
$N=2$ gauged supergravity models coupled to a vector multiplet plus additional
hypermultiplets, and we discuss some explicit examples.

The instabilities will be identified by analysing perturbations of the electrically charged $D=4$ AdS-RN black brane solutions. 
A simple and powerful approach is to first consider linearised perturbations
in the $AdS_2\times \mathbb{R}^2$ background, which arises as the IR limit of the AdS-RN 
geometry at zero temperature.
Depending on the explicit values of various couplings, we find that there are modes which violate the
BF bound for
a range of non-vanishing momentum $k$. This indicates that
there are associated instabilities of the AdS-RN black branes which
set in at a non-zero temperature whose value will depend on $k$.
For the simple model with a single vector field, 
we explicitly construct the static normalisable zero modes and show that there
is a critical momentum $k_c$ which has the highest temperature, $T_c$. At this temperature a new branch of 
striped 
black branes will appear, spontaneously breaking translation invariance with a spatial modulation set, at $T_c$, 
by $k_c$. 
These are dual to a striped 
phase\footnote{The existence of this striped phase obviously requires that the striped black branes appearing at $T=T_c$ 
are thermodynamically preferred.
The thermodynamics of these black branes, as well as the properties of other possible black brane solutions, will be
investigated elsewhere.}
with spatial modulation of the vev of the dual $d=3$ current $(j_t,j_x,j_y)$ given, near $T_c$, by \eqref{eqnone}. As one moves away from $T_c$ higher harmonics will appear, but the spatial modulation of the CDW will have a period that is half
of that of the current density wave.

Before presenting our new results we note that spontaneous breaking of translation invariance has been found in
the context of electrically charged AdS-RN black banes of $D=5$ Einstein-Maxwell theory with a Chern-Simons term in \cite{Nakamura:2009tf}\cite{Ooguri:2010kt}. 
Related earlier work appears in \cite{Domokos:2007kt} and subsequent work appears in \cite{Ooguri:2010xs}\cite{Bayona:2011ab}.
Another\footnote{In \cite{Aperis:2010cd}\cite{Flauger:2010tv} studies were made where translation invariance is
explicitly broken by sources.} holographic investigation of spontaneous breaking of translation invariance in the presence of a
magnetic field was carried out for a $D=4$ Einstein-Yang-Mills-Higgs model in \cite{Bolognesi:2010nb}.

\section{Einstein-Maxwell-pseudo-scalar model}
We consider a class of $D=4$ theories that couples a metric, a gauge field $A$ and a neutral pseudo-scalar $\varphi$
with Lagrangian given by
\begin{align}\label{lagra1}
\mathcal{L}=&\frac{1}{2}R\,\ast 1-\frac{1}{2}\,\ast d\varphi\wedge d\varphi-V\left(\varphi\right) \ast 1- \frac{1}{2}\,\tau\left(\varphi\right) F\wedge\ast F-\frac{1}{2}\,\vartheta\left(\varphi\right) F\wedge F\, ,
\end{align}
where $F=dA$.
The corresponding equations of motion are given by
\begin{align}\label{eomi}
&R_{\mu\nu}=\partial_\mu \varphi\partial_\nu \varphi+g_{\mu\nu}\,V-\tau\,\left(\frac{1}{4}g_{\mu\nu}\,F_{\lambda\rho}F^{\lambda\rho} -F_{\mu\rho}F_{\nu}{}^{\rho}\right)\nn
&d\left(\tau\ast F+\vartheta F\right)=0\nn
&d\ast d\varphi+V'*1+\frac{1}{2}\tau'\,F\wedge\ast F+\frac{1}{2}\vartheta'\,F\wedge F=0\, .
\end{align}
We will assume that the three functions $V$, $\tau$ and $\vartheta$ have the following expansions\footnote{If we ignore the Einstein-Hilbert term in \eqref{lagra1}, 
the $D=4$ models generalise the dimensional reduction of the $D=5$ Maxwell models with a
Chern-Simons term considered in \cite{Nakamura:2009tf}, by the addition of $m_s^2$ and 
$n$. In particular, the tachyonic instability in a uniform electric
field in flat space at finite momentum, considered in section 2 of \cite{Nakamura:2009tf}, has an immediate
extension to $D=4$ massive axions coupled to the electromagnetic field.}
\begin{align}\label{expans}
V=-6+\frac{1}{2}m_{s}^{2}\,\varphi^{2}+\dots,\qquad
\tau=1-\frac{n}{12}\,\varphi^{2}+\dots,\qquad
\vartheta=\frac{c_{1}}{2\sqrt{3}}\,\varphi+\dots\, .
\end{align}
The equations of motion \eqref{eomi} then admit 
the electrically charged $AdS$ Reissner-Nordstr\"om black brane solution
\begin{align}\label{RNbh}
ds^{2}_{4}=& -f\,dt^{2}+\frac{dr^{2}}{f}+r^{2}\,\left(dx^{2}+dy^{2} \right)\nn
A=&\left(1-\frac{r_{+}}{r}\right)\,dt\, ,
%f=&2\,r^{2}-\left(2r_{+}^{2}+\frac{1}{2} \right)\frac{r_{+}}{r}+\frac{r_{+}^{2}}{2r^{2}}
\end{align}
with $\varphi=0$, where
\begin{align}\label{fequals}
f=&2\,r^{2}-\left(2r_{+}^{2}+\frac{1}{2} \right)\frac{r_{+}}{r}+\frac{r_{+}^{2}}{2r^{2}}\, .
\end{align}
In particular, we note that the equation of motion for $\varphi$ in \eqref{eomi} is satisfied with $\vartheta'\ne 0$ because
for the purely electric AdS-RN solution $F\wedge F=0$ (which would not be true for a dyonic AdS-RN black brane).
We also note that we have scaled the chemical potential to be unity, $\mu=1$, for convenience,
and that
the temperature of the black brane is $T=(1/8\pi r_+)(12r_+^2-1)$. 
This solution describes a dual $d=3$ CFT with a global abelian symmetry, whose current, $j$, is dual to $A$,
when held at non-vanishing chemical potential and at high temperatures. Phase transitions can arise if this solution becomes unstable at some critical temperature. We also note 
that $\varphi$ is dual to an operator in the CFT with scaling dimension $\Delta_\pm=(1/2)[3\pm(9+2m_{s}^{2})^{1/2}]$,
with $\Delta_-$ only possible if $-9/2\le m^2_s<-5/2$.

An example of
these models has already appeared in a top-down setting, 
specifically in the context of skew-whiffed $AdS_4\times SE^7$
solutions of $D=11$ supergravity. Recall that these solutions generically do not preserve any supersymmetry, except
in the special case that $SE_7=S^7$ in which case they preserve all of the supersymmetry. 
It was shown in \cite{Gauntlett:2009bh}, building on \cite{Gauntlett:2009zw}\cite{Gauntlett:2009dn}, that there is a 
consistent Kaluza-Klein reduction on an arbitrary $SE_7$ space to a theory involving a metric, a gauge field, a charged 
scalar and a pseudo-scalar. Any solution of this $D=4$ theory can be uplifted to obtain an exact solution
of $D=11$ supergravity. In particular, there is an $AdS_4$ vacuum solution which uplifts to the
skew-whiffed $AdS_4\times SE_7$ solution. It is consistent to further set the charged scalar to zero and the 
resulting $D=4$ model is then as in \eqref{lagra1}, \eqref{expans} with
$n=36$, $c_1= 6\sqrt{2}$ and $m_{s}^{2}=-4$.

\subsection{Perturbative instabilities of $AdS_{2}\times \mathbb{R}^{2}$}\label{AdS2}
The near horizon limit of the extremal ($T=0$) AdS-RN black brane \eqref{RNbh} is the following $AdS_{2}\times \mathbb{R}^{2}$ solution 
\begin{align}\label{ads2sol}
ds_{4}^{2}=&-12r^{2}\,dt^{2}+\frac{dr^{2}}{12r^{2}}+dx^{2}+dy^{2}\nn
F=&2\sqrt{3}\,dr\wedge dt\, ,
\end{align}
with $\varphi=0$ and we have scaled the spatial coordinates by a factor of $2\sqrt 3$ for convenience.
We now examine the following coupled system of perturbations
involving the metric, gauge field and scalar field:
\begin{align}\label{fluctuations}
\delta g_{ty}=&2\sqrt{3}r h_{ty}\left(t,r\right)\sin\left(kx\right)\nn
\delta g_{xy}=&h_{xy}\left(t,r\right)\cos\left(kx\right)\nn
\delta A_{y}=&a\left(t,r\right)\sin\left(kx\right)\nn
\delta \varphi=& w\left(t,r\right)\cos\left(kx\right)\, ,
\end{align}
Note that the subscripts on $h$ are simply to label specific functions of $t$ and $r$.

Substituting into the equations of motion \eqref{eomi} we find that the fluctuations \eqref{fluctuations} satisfy
\begin{align}
2\sqrt{3}k^{2}rh_{ty}-k\partial_{t}h_{xy}-24\sqrt{3}r\,\left(\partial_{r}\left(r^{2}\partial_{r}h_{ty}\right)+2r\partial_{r}a \right)=&0\label{eom1}\\
2\sqrt{3}\,\partial_{t}a+\sqrt{3}\,\partial_{t}h_{ty}+r\,\left(6kr\,\partial_{r}h_{xy}+\sqrt{3}\partial_{r}\partial_{t}h_{ty} \right)=&0\label{eom2}\\
-2\sqrt{3}kr\,\partial_{t}h_{ty}+\partial_{t}^{2}h_{xy}-144r^{2}\partial_{r}\left(r^{2} \partial_{r}h_{xy}\right)=&0\label{eom3}\\
-\frac{1}{12r^2}\partial_{t}^{2}a-k^{2}a+c_{1}kw+12 \left(h_{ty}+\partial_{r}\left(r^{2}\partial_{r}a \right)+r\partial_{r}h_{ty} \right) =&0\label{eom4}\\
-\frac{1}{12r^2}\partial^2_t w+c_{1}ka-\left(m_{s}^{2}+n+k^{2} \right)w+12\,\partial_{r}\left(r^{2}\partial_{r}w \right) =&0\, .\label{eom5}
\end{align}
One can check that equation \eqref{eom3} is implied by equations \eqref{eom1} and \eqref{eom2}. 
Observe that $m_{s}^{2}$ and $n$ only appear in the combination 
\begin{align}
\tilde{m}_{s}^{2}\equiv m_{s}^{2}+n\, .
\end{align}

It is illuminating to now introduce the field redefinition 
\begin{equation}\label{inteqn}
6k\phi_{xy}=-\sqrt{3}\partial_{r}\left(rh_{ty}\right)-2\sqrt{3}a\, .
\end{equation}
We then see that \eqref{eom2} implies
\begin{equation}\label{fredef}
r^{2}\partial_{r}h_{xy}=\partial_{t}\phi_{xy}.
\end{equation}
We also find that
\eqref{eom4}, \eqref{eom5} and the $r$ derivative of \eqref{eom1} can now be packaged in the 
following way. Defining the three vector ${\bf v}=(\phi_{xy},a,w)$
we have
% while from the equations of motion we have the more familiar system
\begin{align}\label{nimr}
\Box_{AdS_2}{\bf v}-M^2{\bf v}=0\, ,
\end{align}
where $\Box_{AdS_{2}}$ is the scalar Laplacian on the $AdS_2$ space with radius squared equal to $1/12$,
and the mass matrix $M^2$ is given by
\begin{equation}\label{mmfirstmod}
M^{2}=\left(\begin{array}{ccc}k^{2} & \frac{1}{\sqrt{3}}k & 0 \\24\sqrt{3}k & 24+k^{2}  & -c_{1}k \\0 & -c_{1}k & k^{2}+\tilde{m}_{s}^{2}\end{array}\right)\, .
\end{equation}

Thus a diagnostic
for an instability is if the mass matrix
has an eigenvalue $m^{2}$ that violates the $AdS_2$ BF bound 
\begin{equation}\label{BFbound}
m^{2}\ge-3\, .
\end{equation}
To be more precise, we have shown that \eqref{nimr}, \eqref{mmfirstmod} are implied by \eqref{eom1}-\eqref{eom5}.
In appendix A we show that, conversely, \eqref{nimr}, \eqref{mmfirstmod} includes all of the perturbations of interest.
Thus, for fixed parameters $\tilde{m}_{s}^{2},c_{1}$ we are looking for ranges of momenta $k$ in which the smallest root
of the characteristic polynomial of $M^{2}$, violates \eqref{BFbound}. In particular, we see that the off-diagonal term in
$M^2$ provides a mechanism, when both $c_1\ne  0$ and $k\ne 0$, to drive down the smallest eigenvalue.

\begin{figure}
\centering
\includegraphics[width=8cm]{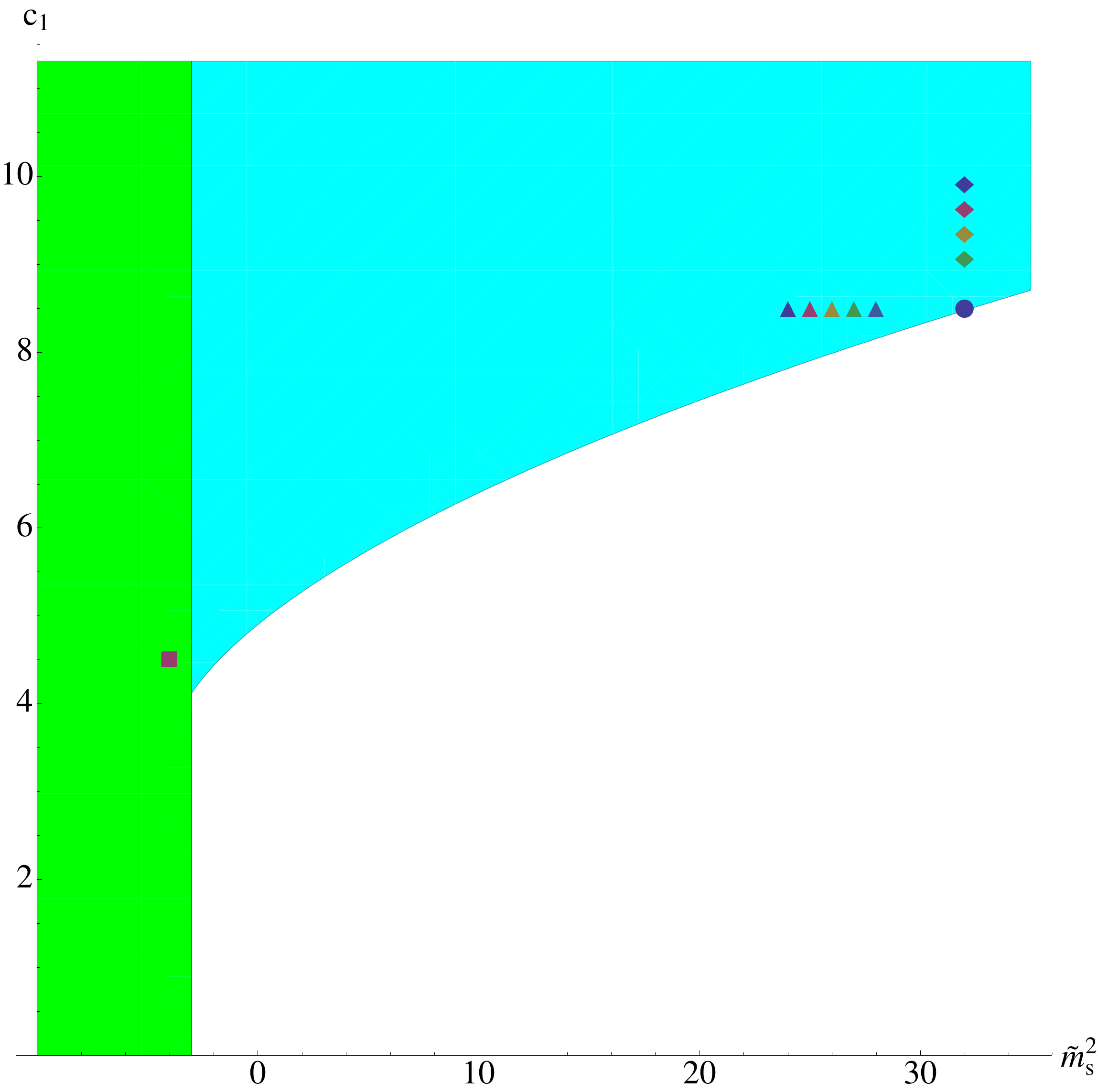}
\caption{The shaded region in the $(\tilde{m}_{s}^{2}, c_{1})$ plane has unstable perturbative modes in the $AdS_{2}\times \mathbb{R}^{2}$
background. The green region has unstable modes including $k=0$ while the cyan region only has unstable modes with 
$k\ne 0$. The blue circle, lying just inside the cyan region,
represents the model associated with skew-whiffed $AdS_4\times SE_7$ solutions. The other symbols are explained in the text.}
\label{fig:theoryMap}
\end{figure}

In figure \ref{fig:theoryMap} we show part of the domain in the $\left(\tilde{m}^{2}_{s},c_{1}\right)$ plane for which there exists a range of momenta %$k_{min}<k<k_{max}$ 
such that one of the corresponding $AdS_{2}$ masses is below the BF bound \eqref{BFbound}. The characteristic polynomial of $M^{2}$ is independent of the sign of $k$ and this leads to two cases. In the first case, marked in green in
figure \ref{fig:theoryMap}, the range of $k$ with unstable modes is of the form $(-k_{max},k_{max})$ 
and in particular contains unstable modes with $k=0$. It is worth noting, though,
 that it is not necessarily the case that the $k=0$ mode is the one with the smallest $AdS_{2}$ mass squared.
In the second case, marked in cyan in figure \ref{fig:theoryMap}, all of the unstable modes have $k\ne 0$ and  
the range of $k$ for which there are unstable modes consists of 
two disjoint regions: $(k_{min},k_{max})$ and its reflection $k\to-k$.

\subsection{Perturbative instabilities of the AdS-RN black brane}
The presence of perturbative instabilities in the $AdS_2\times \mathbb{R}^2$
background that we established in the last subsection suggests that the AdS-RN black brane \eqref{RNbh}, 
\eqref{fequals}
should have analogous perturbative instabilities appearing, for a given $k$, at some specific temperature. 
In particular we shall look for the appearance of static normalisable zero modes which signal the onset of
a dynamical instability. Inspired by
the analysis in the $AdS_{2}\times \mathbb{R}^{2}$ background, we consider the fluctuation
\begin{align}\label{bhfluctuations}
\delta g_{ty}=&\lambda\left[r \left(r-r_{+}\right) h\left(r\right)\sin\left(kx\right)\right]\nn
\delta A_{y}=&\lambda\left[a\left(r\right)\sin\left(kx\right)\right]\nn
\delta \varphi=& \lambda\left[w\left(r\right)\cos\left(kx\right)\right]\, .
\end{align}
Here $\lambda$ is a small expansion parameter and
note that we have not included a time independent perturbation for $g_{xy}$, as suggested by
\eqref{fredef}.  

Expanding around the black brane background \eqref{RNbh} we obtain a linear system of ordinary differential equations which we wish to numerically integrate from the outer horizon, $r=r_{+}$, to asymptotic infinity, $r\to \infty$. In order for the fluctuations \eqref{bhfluctuations} to be regular, the functions $h_{ty}$, $a$ and $w$ should remain finite on the horizon at $r=r_{+}$
\begin{align}\label{pone}
h\left(r\right)=&h_++{\cal O}\,\left(r-r_{+}\right)\nn
a\left(r\right)=&a_++{\cal O}\left(r-r_{+}\right)\nn
w\left(r\right)=&w_++{\cal O}\,\left(r-r_{+}\right)\, .
\end{align}
To see that the metric is regular at the horizon one can use the in-going Eddington-Finkelstein type coordinates $v,r$
where $v\approx t+ln(r-r_+)$.

Unlike in the $AdS_2\times\mathbb{R}^2$ background, the linear ODEs now depend on $m_s$ and $n$ separately. 
For illustration we will focus on the case $m_{s}^{2}=-4$ and quantise so that $\varphi$ corresponds to an operator 
in the dual CFT with scaling dimension $\Delta=2$. By varying $n$ we can still vary $\tilde m_s^2$. 
The most general asymptotic expansion of the functions in \eqref{bhfluctuations} as $r\to \infty$ is
\begin{align}\label{ptwo}
h=&h_{0}+\dots+\frac{h_{3}}{r^{3}}+\ldots\nn
a=&a_{0}+\ldots+\frac{a_{1}}{r}+\ldots\nn
w=&\frac{w_{1}}{r}+\ldots+\frac{w_{2}}{r^{2}}+\ldots\, .
\end{align}
The parameters $h_0$, $a_0$ and  $w_1$ correspond to deforming the dual field theory by the operators dual to the fields, while
$h_3$, $a_1$ and $w_2$ correspond to the operators acquiring vev's. We are interested in looking for instabilities that spontaneously break
translational invariance so we set $h_{0}=a_{0}=w_1=0$. 

We have three second order linear ODE's to solve and so a solution is specified by six integration constants.
Now, for a given $k$, we have seven parameters $r_+, h_+,a_+,w_+,h_3,a_1,w_2$ entering the ODEs (since $\lambda$ drops out). 
However, since the equations are linear we can always scale one of these parameters to unity. Hence, for a given $k$ we 
expect a normalisable zero mode to appear, if at all, at a specific temperature.

\begin{figure}
\centering
\subfloat[$n=0$, $c_1= 4.5$]{\includegraphics[width=5cm]{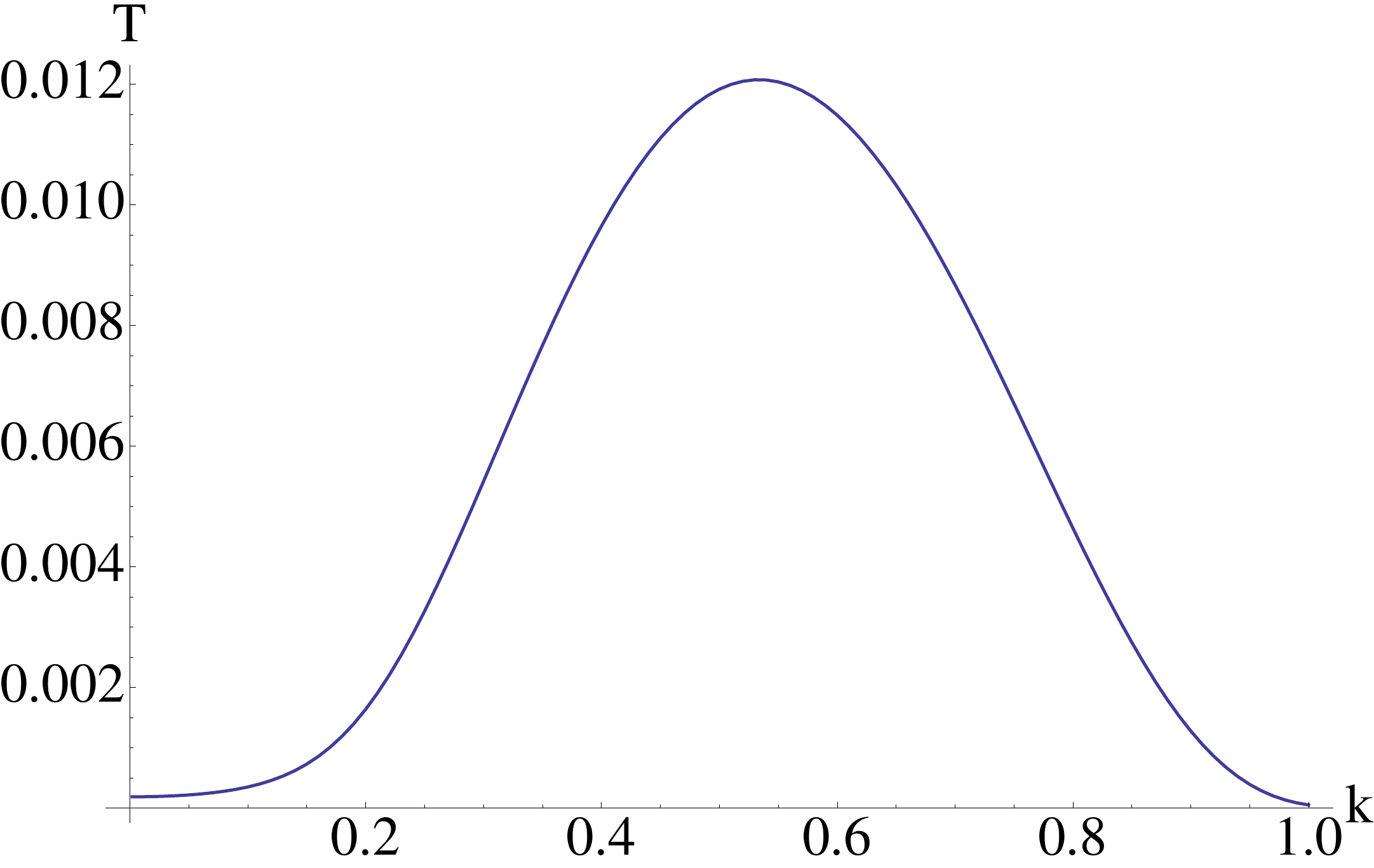}\label{fig:a}}
\subfloat[$n=36$, various $c_1$]{\includegraphics[width=5cm]{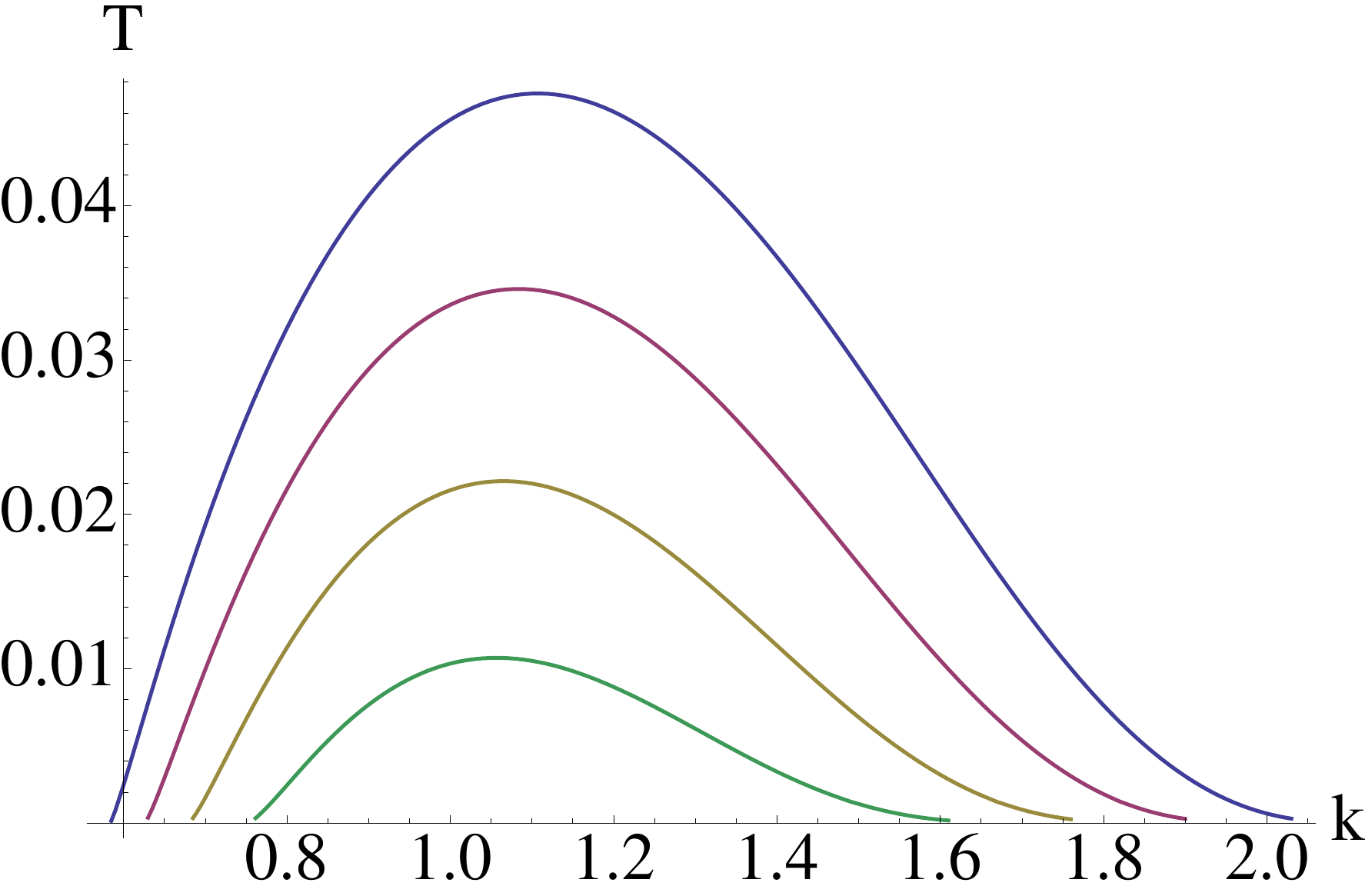}\label{fig:b}}
\subfloat[$c_{1}=6\sqrt{2}$, various $n$]{\includegraphics[width=5cm]{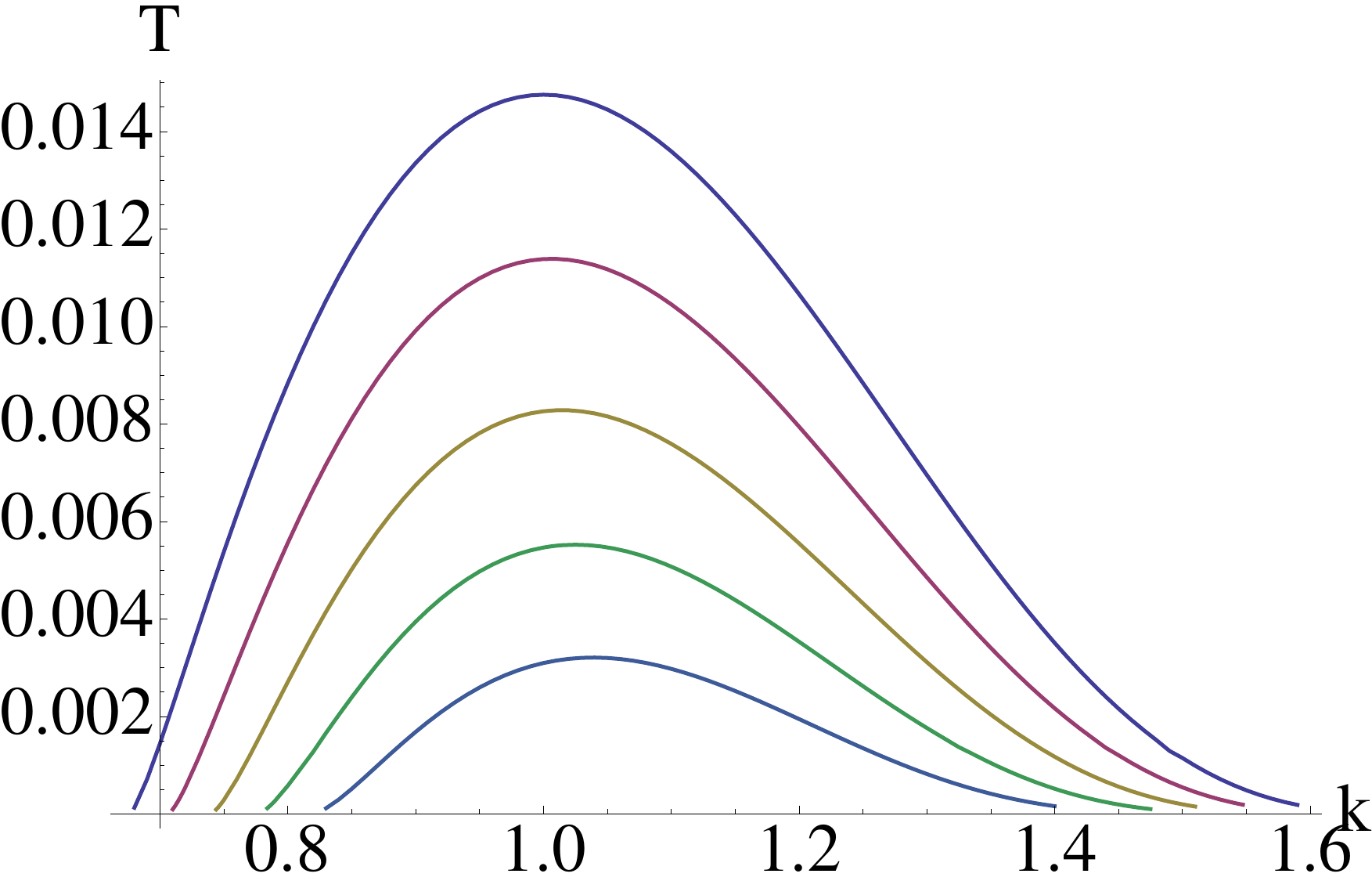}\label{fig:c}}
\caption{Plots of critical temperatures $T$ versus $k$ for the existence of normalisable static perturbations
about the electrically charged 
AdS-RN black brane. All cases have $m^2_s=-4$. With reference to figure \ref{fig:theoryMap}, 
figure (a) corresponds to the square, figure (b) corresponds to the rhombi and
figure (c) corresponds to the triangles.}
\label{fig:charged_VEVs_single}
\end{figure}

We have numerically studied the existence of normalisable modes for various specific values of $n$ and $c_1$.
We begin with a case with $n=0$. We choose $c_{1}= 4.5$ to illustrate the affect of turning on $c_1$. Note that
this case is marked by the square in the green area in figure \ref{fig:theoryMap}.
When $k=0$ (and hence $c_1k=0$), this case was already analysed in \cite{Denef:2009tp} 
($\Delta=2$ and $q=0$ in their language). There it was shown that normalisable perturbative modes with $k=0$ are present with critical temperature $T<10^{-3}$. 
Here we find that perturbative static normalisable zero modes also exist for a range of $k$, with the critical temperatures depending on $k$ as shown 
in figure \ref{fig:a}. 
%We see that there is indeed a static mode with $k=0$ with critical  
%temperature $T<10^{-3}$ in agreement with the results of figure 1 in \cite{Denef:2009tp}.  
The highest critical temperature occurs for wavenumber $k\approx 0.53$ and has $T_c\approx 0.012$.
At this temperature, a new branch of black branes will appear that spontaneously break translation symmetry. 

We next consider the blue dot in figure \ref{fig:theoryMap}, corresponding to the top-down models associated with
the skew-whiffed $AdS_4\times SE_7$ solutions. The maximum critical temperature $T_c$ is very low for this case, so we 
illustrate what is going on by studying the sequence of values of $n$ and $c_1$ given by the rhombi and triangles in
figure \ref{fig:theoryMap}. Plots of the critical temperatures for different $k$ for which there is a static perturbative 
normalisable mode are given in figures \ref{fig:b} and \ref{fig:c}. For all these cases, we find that the range of momenta
$k$ for which there is an unstable static zero mode does not include $k=0$ as expected from the
the analysis in the $AdS_2\times \mathbb{R}^2$ background. The figures indicate that the critical temperature
for the skew-whiffed case is going to be very low, but certainly for a non-zero value of $k$. It is also worth noting that the maximum critical temperature is
much lower than the critical temperature for the known superfluid instability which is given by $T\approx 0.042$.

\subsection{Stripes}
In the last subsection we constructed the normalisable zero modes in the AdS-RN black brane background at leading order in perturbation theory. There is a critical value of momentum $k_c$ which is associated with
the highest critical temperature $T_c$ at which the zero modes appear (the maxima in figure 
\eqref{fig:charged_VEVs_single}). When $T=T_c$ a new branch of black branes will appear that spontaneously
break translation invariance. In the leading order perturbations given in \eqref{pone} and \eqref{ptwo} we find
that $w_2$, $h_3$ and $a_1$ are all non-zero, as the simple counting of integration constants above indicated. 
$w_2\ne 0$ implies that 
the scalar operator dual to $\varphi$ has acquired a 
spatially modulated vev. 
Similarly, $h_3\ne 0$ implies that there is momentum transfer in the $x^2$ direction. Finally, $a_1\ne 0$ implies
that the dual current component $\langle j_y\rangle$ is also acquiring a spatially modulated vev of the form \eqref{eqnone}. 
Thus, the new black branes are dual to a current density wave phase with spatial modulation given, at $T_c$, 
by $k_c$. 
As we will now argue the phase is also a charge density wave (CDW) with $\langle j_t\rangle$ 
of the form \eqref{eqnone} and hence, near $T_c$, a spatial modulation given by $2k_c$. 

To see the CDW, we need to analyse the general structure of the equations arising in the next to leading
order in perturbation theory. One can show that a closed system of equations is obtained if we take the second order
perturbations to be given by
\begin{align}\label{eq:2nd_expansion}
\delta g_{tt}&=\lambda^{2} \left[ h_{tt}^{(0)}\left(r\right)+h_{tt}^{(1)}\left(r\right)\,\cos\left(2kx \right) \right]\nn
\delta g_{xx}&=\lambda^{2} \left[ h_{xx}^{(0)}\left(r\right)+h_{xx}^{(1)}\left(r\right) \,\cos\left(2kx \right)\right]\nn
\delta g_{yy}&=\lambda^{2} \left[ h_{yy}^{(0)}\left(r\right)+h_{yy}^{(1)}\left(r\right) \,\cos\left(2kx \right)\right]\nn
\delta A_{t}&=\lambda^{2} \left[ a_{t}^{(0)}\left(r\right)+a_{t}^{(1)}\left(r\right)\,\cos\left(2kx \right) \right]\, .
%\delta B_{t}&=\lambda^{2} \left( B_{t}^{(0)}\left(r\right)+B_{t}^{(1)}\left(r\right)\,\cos\left(2kx \right) \right)
\end{align}
Plugging the total field expansion in the equations of motion \eqref{eomi} and expanding them up to order $O(\lambda^{2})$ we obtain an inhomogeneous system of ordinary differential equations, being sourced
by the $O(\lambda)$ zero mode solution. 
More specifically, the functions $h_{xx}^{(\alpha)}$, $h_{yy}^{(\alpha)}$ and $a_{t}^{(\alpha)}$ satisfy second order equations, the function $h_{tt}^{(0)}$ satisfies a first order equation while $h_{tt}^{(1)}$ satisfies an algebraic equation and hence
can be eliminated from the system.

We next need to impose regularity at the horizon and demand that the behaviour as $r\to\infty$ corresponds 
to setting all source terms in the dual CFT to zero.
A simple count of parameters and integration constants then indicate 
that for a given $k$, and in particular for $k=k_c$, the zero mode found in the last subsection forms
part of a one-parameter branch of spatially modulated black brane solutions, where the parameter can be taken to be the temperature. We will expand on the solutions to these
ODEs in more detail elsewhere, but the main point we wish to emphasise here is that, generically, the asymptotic falloff 
of $a^{(1)}_t(r)$ as $r\to\infty$, in particular, will be of the form
\begin{align}
a^{(1)}_t(r)={0}+\frac{\bar a^{(1)}_t}{r}+\dots
\end{align}
implying that there is a spontaneous spatial modulation of the charge density with, from \eqref{eq:2nd_expansion},
characteristic wavenumber given by $2k_c$, at $T_c$, 
and in the same direction as the current density wave. Note that $a_{t}^{(0)}$ in 
\eqref{eq:2nd_expansion} will have a similar asymptotic behaviour and the non-vanishing $1/r$ component will
give rise to the $\bar j_t$ piece in \eqref{eqnone}.
Thus the new branches
of black brane solutions, assuming that they are thermodynamically preferred,
will be dual to striped phases
incorporating both current density waves and CDWs.

Note that the perturbative expansion parameter $\lambda$ can be taken to be $(T-T_c)/T_c$.
For small $\lambda$, i.e. near $T_c$, 
we have argued that the wavenumber for the spatial modulation of the current density wave will be
$k_c$ and for the CDW will
be $2k_c$. Following an analogous discussion in \cite{Gubser:2001ac}, continuing to higher orders in the
perturbative expansion we expect that the wavenumber will receive corrections at order $\lambda^2$. However, the spatial
modulation of the CDW will have a period that is always half that of the current density wave and in the same direction.

\section{A pseudo-scalar coupled to two vector fields}
We now consider a more general class of $D=4$ theories that couple a metric, a pseudo-scalar and 
two vector fields with Lagrangian given by
\begin{align}\label{laggen}
\mathcal{L}=\frac{1}{2}R\,\ast 1&-\frac{1}{2}\ast d\varphi \wedge d\varphi-V\left(\varphi\right)\,\ast 1-\frac{1}{2}\,\tau\left(\varphi\right)\,F\wedge\ast F -\frac{1}{2}\,\vartheta\left(\varphi\right)\,F\wedge F\nn
&-\frac{1}{2}G\wedge \ast G-\frac{1}{2}m_{v}^{2}\ast B\wedge B+\frac{c_2\varphi}{2\sqrt{3}}\,F\wedge G\, ,
\end{align}
where $F=dA$ and $G=dB$. In the first line we take
\begin{align}
V=-6+\frac{m_s^2}{2}\varphi^2+\dots,\qquad
\tau=1-\frac{n}{12}\varphi^2+\dots,\qquad
\vartheta=\frac{c_1}{2\sqrt 3}\varphi+\dots\, ,
\end{align}
as in the last section, while in the second line we have introduced two new parameters $m^2_v$ and $c_2$.
In the $AdS_4$ vacuum with $A=B=\varphi=0$, the gauge field $A$ is massless, while $B$
has mass $m_v$. These models admit the electrically charged AdS-RN
black brane solution, \eqref{RNbh}, \eqref{fequals} with $B=\varphi=0$, as a solution. This solution describes a
dual $d=3$ CFT at high temperature and finite chemical potential with respect to the global abelian symmetry
whose current is dual to the gauge field $A$. The scalar field, $\varphi$, is dual to a scalar operator with conformal
dimension given before and the second vector field, $B$, is dual to a vector operator with conformal dimension 
$\Delta=(1/2)[3+(1+2m^2_v)^{1/2}]$.
%At zero temperature and in the near horizon limit 
%this approaches the $AdS_2\times \mathbb{R}^2$ solution \eqref{ads2sol}. 

We are interested in perturbative instabilities of the electrically charged AdS-RN black brane solution. 
We first point out that \eqref{laggen} can be generalised in a number of obvious ways, including
adding a $f(\varphi) G \wedge G$ term, without affecting the linearised analysis that we cary out below. 
Furthermore, we can also couple additional matter fields.
Thus the instabilities that we discuss for \eqref{laggen} will capture a large class of examples arising in string and M-theory.
For example, it overlaps with the following two cases that have been explicitly considered in the literature
that involve $N=2$ gauged supergravity coupled to a vector multiplet plus additional hypermultiplets.

The first case is the
consistent truncation of $D=11$ supergravity on a $SE_7$ to a $D=4$ theory whose $AdS_4$ vacuum uplifts
to the supersymmetric $AdS_4\times SE_7$ solution of $D=11$ supergravity \cite{Gauntlett:2009zw}. The matter content of
the $D=4$ theory consists of a metric, two vectors and six scalars, which package together into an $N=2$ gravity multiplet,
a vector multiplet and a hypermultiplet, and in particular the vector multiplet contains a pseudo-scalar labelled as $h$ in 
\cite{Gauntlett:2009zw}. We find that the linearised perturbations for \eqref{laggen} that we consider below 
arise in a sector of the $D=4$ theory of \cite{Gauntlett:2009zw}. To see this one should identify the 
field strengths via $H^{there}=1/2(F-G/\sqrt{3})$, $F^{there}=1/2(F+\sqrt{3} G)$, the scalar $h=-(\sqrt{2}/\sqrt{3})\varphi$
and finally rescale the metric $g^{there}_{\mu\nu}=1/2 g_{\mu\nu}$. One then finds that there are linearised
perturbations involving the pseudo-scalar which are exactly the same as those coming from \eqref{laggen} with
\begin{align}\label{susyparam}
m^2_s=20,\quad m^2_v=24,\quad n=12,\quad c_1=0,\quad c_2=2\sqrt{6}\, .
\end{align}

The second case is the consistent KK reduction of $D=11$ supergravity on 
$H_3\times S^4$ where $H^3$ is three-dimensional hyperbolic space \cite{Donos:2010ax} (in fact $H^3$ can be replaced by an arbitrary quotient $H^3/\Gamma$). 
The resulting $D=4$ theory
has an $AdS_4$ vacuum solution which uplifts to a supersymmetric $AdS_4\times H^3\times S^4$ solution of $D=11$
supergravity and is dual to a $d=3$ $N=2$ SCFT that arises on M5-branes wrapping special Lagrangian
3-cycles $H^3$ \cite{Gauntlett:2000ng}. The matter content of
this $D=4$ theory consists of a metric, two vectors and ten scalars, which package together into an $N=2$ gravity multiplet,
a vector multiplet and two hypermultiplets, and in particular the vector multiplet contains a pseudo-scalar labelled as $\beta$ in 
\cite{Gauntlett:2000ng}. We find that the linearised perturbations for \eqref{laggen} that we consider below also
arise in a sector of the $D=4$ theory of \cite{Gauntlett:2000ng}. To see this one should identify the 
field strengths via $\tilde H^{there}=2^{-3/4}(F+G/\sqrt{3})$, $F^{there}=2^{1/4}(F-G/\sqrt{3})$, the scalar $\beta=(1/\sqrt{3})\varphi$
and finally choose the gauge coupling $g^{there}=2^{3/4}$. One then finds that there are linearised
perturbations involving the pseudo-scalar which are exactly the same as those coming from \eqref{laggen} with
\begin{align}\label{h3s4}
m^2_s=4,\quad m^2_v=8,\quad n=12,\quad c_1=0,\quad c_2=2\sqrt{6}
\end{align}

Notice that both of these examples have $c_1=0$. This was actually to be expected for these supersymmetric examples. 
First observe that $c_1=0$ is required in order to be able to consistently truncate the theory \eqref{laggen}
to the Einstein-Maxwell sector involving the metric and the gauge field $A$
(otherwise there would be a $F\wedge F$ source term in the equation of motion for $\varphi$). Second, we recall that
for any $AdS_4\times M$ solution of $D=10$ or $D=11$ supergravity which is dual to an $N=2$ SCFT in $d=3$, 
it is conjectured that there is a consistent Kaluza-Klein truncation on $M$ to $N=2$ $D=4$ minimal gauged supergravity, and this
has been proven for several different classes \cite{Gauntlett:2007ma}. The relevant gauge field is dual to a canonical ``Reeb" Killing vector of $M$.
Since the bosonic sector of minimal gauged supergravity is simply Einstein-Maxwell theory, the more general $N=2$ supergravity theories with $A$
the canonical gauge field
must have\footnote{Recall that in section 2 we argued that associated with 
the skew-whiffed $AdS_4\times SE_7$ solutions, which generically don't preserve
supersymmetry, there is a truncation with $c_1=6\sqrt{2}$. For the special case when $SE_7=S^7$ the models preserve $N=8$ supersymmetry.
This is consistent with the discussion here because the gauge field being kept is not a canonical gauge field
dual to a Reeb vector compatible with the orientation of the $S^7$.} 
 $c_1=0$. 

We now investigate instabilities of the electrically charged AdS-RN black brane solution of \eqref{laggen} by considering the
following linearised perturbations in the $AdS_{2}\times \mathbb{R}^{2}$ background:
\begin{align}
\delta g_{ty}=&2\sqrt{3}r\,h_{ty}\left(t,r\right)\,\sin\left(kx\right)\nn
\delta g_{xy}=&h_{xy}\left(t,r\right)\,\cos\left(kx\right)\nn
\delta A_{y}=&a\left(t,r \right)\,\sin\left(kx\right)\nn
\delta B_{y}=&b\left(t,r\right)\sin\left(k x\right)\nn
\delta\varphi= &w\left(t,r\right)\,\cos\left(kx\right)\, .
\end{align}
After substituting into the equations of motion arising from \eqref{laggen} we are lead to a system of 
six linear equations, one of which is implied by the others.
It is again useful to now introduce the field redefinition as in \eqref{inteqn}.
%\begin{equation}\label{fredef}
%$r^{2}\partial_{r}h_{xy}=\partial_{t}\phi_{xy}$ 
%and once more we find that
%one of the equations can be integrated exactly
%as in \eqref{inteqn}.
%\end{equation}
Then, similar to before, after defining the four vector ${\bf v}=(\phi_{xy},a,w,b)$, the remaining independent equations 
imply that
%the more familiar looking equations
% while from the equations of motion we have the more familiar system
\begin{align}
\Box_{AdS_2}{\bf v}-M^2{\bf v}=0\, ,
\end{align}
where $\Box_{AdS_{2}}$ is the scalar Laplacian on the $AdS_2$ space with radius squared equal to $1/12$,
and the mass matrix $M^2$ is given by
\begin{equation}\label{mmsecondmod}
M^{2}=\left(\begin{array}{cccc}k^{2} & \frac{1}{\sqrt{3}}k & 0 &0 \\24\sqrt{3}k & 24+k^{2}  & -c_{1}k &0 \\0 & -c_{1}k & \tilde m_{s}^{2} +k^2& c_{2}k\\ 0 &0 & c_{2}k& m_{v}^{2}+k^{2}\end{array}\right)
\end{equation}
and $\tilde m^2_s\equiv m^2_s+n$.

This mass matrix can have eigenvalues violating the $AdS_2$ BF bound \eqref{BFbound} for non-zero values of $k$. 
The key features are the off-diagonal entries $c_1k$ and $c_2k$.
The general analysis is not that illuminating so we shall not present it here. Instead we make
the following observations. We first notice that when $c_2=0$, the second vector field, $B$, decouples at the linearised level. Indeed when $c_2=0$ the upper $3\times 3$ block is exactly the same
as we found in \eqref{mmfirstmod} in the last section.

We next consider the special case with $c_1=0$, which in particular arises in the
KK truncations of $D=11$ supergravity associated with the supersymmetric $AdS_4\times H^3\times S^4$ and  \
$AdS_4\times SE_7$ solutions. 
Notice that the mass matrix \eqref{mmsecondmod}
is now block diagonal
and that possible BF violating modes must appear in the lower $2\times 2$ block. In particular, we see
that the gravitational perturbations have decoupled from this sector\footnote{In the context of $AdS\times sphere$ solutions instabilities at finite momentum that are decoupled
from gravitational perturbations have been studied in \cite{Lu:2010aua}.}. The eigenvalues of this block are given
by
\begin{equation}
m^{2}_{\pm}=\frac{1}{2}\,\left(2k^{2}+\tilde{m}_{s}^{2}+m_{v}^{2}\pm\sqrt{(2c_2k)^{2}+\left(\tilde{m}_{s}^{2}-m_{v}^{2} \right)^{2}} \right)\, .
\end{equation}
One can easily show that for $c_2^{4}>\left(\tilde{m}_{s}^{2}-m_{v}^{2} \right)^{2}$ the branch $m_{-}^{2}$ develops a minimum at a non-zero value of $k$ given by
\begin{align}
k=\pm\frac{\sqrt{c_2^4-\left(\tilde{m}_{s}^{2}-m_{v}^{2} \right)^{2}}}{2c_2}\, ,
\end{align}
with the minimum $AdS_{2}$ mass given by
\begin{equation}
m^{2}_{min}=-\frac{1}{4c_2^{2}}\,\left[c_2^{4}+\left(\tilde{m}_{s}^{2}-m_{v}^{2} \right)^{2}-2c_2^{2}\,\left(\tilde{m}_{s}^{2}+m_{v}^{2} \right) \right]\, .
\end{equation}
We now see that for sufficiently large $c_2$ one can always have $m_{min}^{2}<-3$ violating the BF bound \eqref{BFbound}.
For the explicit parameters arising in the KK truncations associated with the supersymmetric $AdS_4\times SE_7$ and $AdS_4\times H^3\times S^4$ solutions that were given in \eqref{susyparam} and \eqref{h3s4}, respectively, we see that these perturbations involving the pseudo-scalar
do not violate the BF bound. It would be interesting to know whether or not there are supersymmetric top down models with $c_1=0$
and large enough $c_2$ to give instabilities.

Returning to the general analysis, as in the last section, we find that the instabilities in the $AdS_2\times \mathbb{R}^2$ background 
are associated with static normalisable zero modes appearing in the AdS-RN black brane at, for a given $k$,
some critical temperature. We have investigated some specific examples and the results are similar
to those presented in figure \eqref{fig:charged_VEVs_single}.

\section{Discussion}
In this paper we have identified new instabilities of $D=4$ electrically charged AdS-RN black branes
in a broad class of models involving pseudo-scalars. Generically, the instabilities appear at non-vanishing spatial
momentum. The static normalisable modes that we identified correspond to the appearance of new
branches of striped black brane solutions with the spatial modulation of the CDW being half that of the 
current density wave and in the same direction. It will be interesting to investigate the thermodynamics of the striped black branes and establish the conditions for which they are thermodynamically preferred over the AdS-RN black branes. 
We expect at least in some cases that the phase transitions are second order and we will be able 
to verify this by developing the perturbative expansion we have used in this paper. However, if they are first order 
we would need 
to construct the fully back reacted branes by solving PDEs. This will also be necessary to be able to
follow the striped phases to their ultimate zero temperature ground states. Alternatively, perhaps it
is possible to construct candidate zero temperature ground states directly.

We showed that spatially modulated instabilities are present in a $D=4$ model associated with the skew-whiffed
$AdS_4\times SE_7$ solutions of $D=11$ supergravity. We showed that the maximal critical temperature at which
the instabilities appear is lower than the critical temperature associated with the superfluid instability. 
However, it seems likely that the two instabilities will compete, possibly after a deformation by a relevant operator, and will
lead to thermodynamically preferred striped phases which may also be superconducting. We find this a particularly
interesting direction to pursue given the potential similarities with what is seen in the heavy fermion and
the high temperature superconductors.

\subsection*{Acknowledgements}
We would like to thank Fay Dowker, Sung-Sik Lee, Subir Sachdev, Toby Wiseman and Jan Zaanen
for helpful discussions. 
AD is supported by an EPSRC Postdoctoral Fellowship.
JPG is supported by an EPSRC Senior Fellowship and a Royal Society Wolfson Award.

\appendix

\section{A converse result}
We showed that \eqref{eom1}-\eqref{eom5} imply \eqref{nimr}, \eqref{mmfirstmod}. Let us now
prove a converse result. We start with a regular perturbation in the $AdS_2\times\mathbb{R}^2$
background satisfying \eqref{nimr}, \eqref{mmfirstmod}. We can then integrate 
\eqref{fredef} and \eqref{inteqn} to define $r\,h_{ty}$ and $h_{xy}$ up to two
arbitrary integration functions of time, $z_1(t)$ and $z_2(t)$. Note that \eqref{eom2} is automatically satisfied.
On the other hand, we find that demanding that \eqref{eom1} is satisfied, and furthermore
demanding that $r\,h_{ty}=0$ at $r=0$ (i.e. the perturbation \eqref{fluctuations} is regular at $r=0$), fixes $z_1$ and $z_2$. Indeed we
find that
\begin{align}\label{eq:ht2} 
r\,h_{ty}&=-\,\int_{0}^{r}dr\left(2a+2\sqrt{3}k\,\varphi_{xy}\right)\, ,\nn
h_{xy}&=\int\,dt \left[144r^{2}\,\partial_{r}\varphi_{xy}-\int_{0}^{r} dr\left(4\sqrt{3}k\,a+12k^2\,\varphi_{xy} \right) \right]\, .
\end{align}

In order to make the asymptotic behaviour of \eqref{eq:ht2} more transparent we will use the three normal modes, $s_{\alpha},\,\alpha=1,2,3$, coming from diagonalising the system of equations \eqref{nimr}. Specifically, 
if the eigenvalues of the mass matrix \eqref{mmfirstmod} are $m_{\alpha}^{2}$ we have
\begin{equation}\label{eq:box_equation}
\Box_{AdS_{2}}s_{\alpha}-m_{\alpha}^{2}\,s_{\alpha}=0
\end{equation}
We can then rewrite \eqref{eq:ht2} as
\begin{align}
r\,h_{ty}&=-2\sqrt{3}k\,\sum_{\alpha}c_{\alpha}\,\int_{0}^{r}dr\,m_{\alpha}^{2}s_{\alpha}\label{eq:ht2a}\nn
h_{xy}&=12\,\sum_{\alpha}c_{\alpha}\,\int\,dt\,\left[\frac{1}{12}r^{2}\partial_{r}s_{\alpha}-\int_{0}^{r}dr\,m_{\alpha}^{2}s_{\alpha} \right]
\end{align}
for some constants $c_{\alpha}$. We now focus on a specific mode, $s$, and assume a time dependence of the form $e^{-i\omega t}$. The 
normalisable modes of \eqref{eq:box_equation} will behave as
\begin{equation}
s\approx \frac{1}{r^{\delta(k)}}
\end{equation}
with $\delta(k)>1/2$ for non-tachyonic modes and $Re(\delta(k))=1/2$ for tachyonic modes.
Using explicit expressions for $s$ in terms of Bessel functions
% we can show that in fact the integral over the radial coordinate in \eqref{eq:ht2a} and \eqref{eq:h12a} is not divergent and that asymptotically 
we have the schematic expansions
\begin{align}\label{eq:asympt}
rh_{ty}&\approx e^{-i\omega t}\left[ -\frac{2\sqrt{3}}{k} A+ \frac{f_{1}}{r^{\delta(k)-1}}\right]\nn
h_{xy}&\approx e^{-i\omega t}\left[-\frac{12i A}{\omega}+\frac{f_{2}}{r^{\delta(k)+1}}\right]
\end{align}
where $A$ is a finite constant.
Finally, we point out that the dependence on $A$ in \eqref{eq:asympt} is actually a gauge artifact
and hence is not modifying the $AdS_{2}\times \mathbb{R}^{2}$ asymptotics. 
Indeed it can be eliminated by performing the (perturbative) coordinate change
\begin{equation}
y\rightarrow y+\frac{12i}{k\omega}e^{-i\omega t} A\,Q\left(r\right)\, \sin\left(kx \right)
\end{equation}
with $Q$ a sufficiently smooth function such that the first few derivatives vanish at $r=0$ and $r=\infty$ and also $Q(0)=0$, $Q(\infty)=1$. 
Note that this change of coordinates also introduces a non-zero $\delta g_{ry}$ perturbation, but it has support only in the bulk and vanishes at both
the origin and boundary and is thus benign.

\bibliographystyle{utphys}
\bibliography{AxionInst}{}
\end{document}